\documentstyle[12pt]{article}
\begin{document}
\begin{center}{\large{\bf Physical Principles Based on Geometric Properties}}
\end{center}
\vspace*{1.5cm}
\begin{center}
A. C. V. V. de Siqueira
$^{*}$ \\
Departamento de Educa\c{c}\~ao\\
Universidade Federal Rural de Pernambuco \\
52.171-900, Recife, PE, Brazil.\\
\end{center}
\vspace*{1.5cm}
\begin{center}{\bf Abstract}

In this paper we present some results obtained in a previous paper
about the Cartan's approach to Riemannian normal coordinates and our
conformal transformations among  pseudo-Riemannian manifolds. We
also review the classical and the quantum angular momenta of a
particle obtained as a consequence of geometry, without postulates.
We present four classical principles, identified as new results
obtained from geometry. One of them has properties similar to the
Heisemberg's uncertainty principle and another has some properties
similar to the Bohr's principle. Our geometric result can be
considered as a possible starting point toward a quantum theory
without forces.
  \end{center}

 \vspace{3cm}

${}^*$ E-mail: acvvs@ded.ufrpe.br
\newline

\newpage

\section{Introduction}
$         $

In a previous paper dedicated only to mathematical results
\cite{1},we show how n-dimensional pseudo-Riemannian spaces are
related to each other by a conformal transformation. As a geometric
consequence, we obtained the classical angular momentum and the
quantum angular momentum operator of a particle, regardless of a
physical theory. In this paper, suggested by geometric results, we
build some new principles and consider the possibility of a new
starting point toward a quantum theory without forces.
 \newline
This paper is organized as follows. In Sec.$2$ we present normal
coordinates and elements of differential geometry. In Sec.$3$ we
show that all well-behaved n-dimensional pseudo-Riemannian metrics
in origin and in its neighborhood, in normal coordinates, are
conformal to an n-dimensional flat manifold and to an n-dimensional
manifold of constant curvature. In Sec.$4$, we make an embedding of
all n-dimensional pseudo-Riemannian manifold of constant curvature
in an n+1-dimensional flat manifold, obtaining, without postulates,
the quantum angular momentum operator of a particle as  a
consequence of geometry. In Sec.$5$, based on geometric properties,
we present some physical principles. Section $6$ is a continuation
of section $5$ with more concentration in quantum mechanics.
\renewcommand{\theequation}{\thesection.\arabic{equation}}
\newpage
\section{\bf  Normal Coordinates}
$         $
 \setcounter{equation}{0}
 $         $
 \setcounter{equation}{0}
 $         $
In this section we briefly present normal coordinates and  review
some elements of differential geometry for an n-dimensional
pseudo-Riemannian  manifold, \cite{2}, \cite{3}, \cite{4}.
\newline
Let us consider the line element
\begin{equation}
 ds^2= G_{\Lambda\Pi}du^{\Lambda}du^{\Pi},
\end{equation}
with
\begin{equation}
G_{\Lambda\Pi}=E_{\Lambda}^{(\mathbf{A})}E_{\Pi}^{(\mathbf{B})}\eta_{(\mathbf{A})(\mathbf{B})},
\end{equation}
where $ \eta_{(\mathbf{A})(\mathbf{B})}$ and $
E_{\Lambda}^{(\mathbf{A})}$ are flat metric and vielbein
components respectively.
\newline
We choose each
 $ \eta_{(\mathbf{A})(\mathbf{B})}$ as a plus or minus  Kronecker's  delta function.

Let us give the 1-form  $\omega^{(\mathbf{A})} $ by
\begin{equation}
\omega^{(\mathbf{A})}= du^{\Lambda} E_{\Lambda}^{(\mathbf{A})}.
\end{equation}
We now define Riemannian normal coordinates by
\begin{equation}
  u^{\Lambda}=v^{\Lambda}t,
\end{equation}
Substituting in (2.3)
\begin{equation}
  \omega^{(\mathbf{A})}= tdv^{\Lambda}
  E_{\Lambda}^{(\mathbf{A})}+dtv^{\Lambda}E_{\Lambda}^{(\mathbf{A})}.
\end{equation}
Let us define
\begin{equation}
 z^{(\mathbf{A})}=v^{\Lambda}E_{\Lambda}^{(\mathbf{A})},
\end{equation}
so that
\begin{equation}
\omega^{(\mathbf{A})}=dtz^{(\mathbf{A})}+tdz^{(\mathbf{A})}
+tE^{\Pi(\mathbf{A})}\frac{\partial{E_{\Pi(\mathbf{B})}}}{\partial{z^{(\mathbf{C})}}}z^{(\mathbf{B})}dz^{(\mathbf{C})}.\\
\end{equation}
We now make
\begin{equation}
 A^{(\mathbf{A})_{(\mathbf{B})(\mathbf{C})}}=tE^{\Pi(\mathbf{A})}\frac{\partial{E_{\Pi(\mathbf{B})}}}{\partial{z^{(\mathbf{C})}}},
\end{equation}
then
\begin{equation}
\varpi^{(\mathbf{A})}=
tdz^{(\mathbf{A})}+A^{({A})_{(\mathbf{B})(\mathbf{C})}}z^{(\mathbf{B})}dz^{(\mathbf{C})},
\end{equation}
with
\begin{equation}
\omega^{(\mathbf{A})}=dtz^{(\mathbf{A})}+\varpi^{(\mathbf{A})}.
\end{equation}
We have at $t=0$
\begin{equation}
 A^{({A})_{(\mathbf{B})(\mathbf{C})}}(t=0,z^{(\mathbf{D})})=0,
\end{equation}
\begin{equation}
\varpi^{(\mathbf{A})}(t=0,z^{(\mathbf{D})})=0 ,
\end{equation}
and
\begin{equation}
\omega^{(\mathbf{A})}(t=0,z^{(\mathbf{D})})=dtz^{(\mathbf{A})} .
\end{equation}
Consider, at an n+1-manifold, a coordinate system given by
$(t,z^{(\mathbf{A})})$. For each value of t we have a hyper-surface,
where $dt=0$ on each of them. We are interested in the hyper-surface
with $t=1$, where we verify the following equality
\begin{equation}
\omega^{(\mathbf{A})}(t=1,z)=\varpi^{(\mathbf{A})}(t=1,z).
\end{equation}
\newline
It is well known the following expression in a  vielbein basis
\begin{equation}
d\omega^{(\mathbf{A})}=-\omega^{(\mathbf{A})}_{(\mathbf{B})}\wedge\omega^{(\mathbf{B})}.
\end{equation}
Considering now the map  $\Phi$, between two manifolds M and N,
\newline
and two subsets, U of M and V of N, we have
\begin{equation}
\Phi:U\longrightarrow V.
\end{equation}
Defining now pull-back as follows, \cite{3}, \cite{4},
\begin{equation}
\Phi^\ast:F^p(V)\longrightarrow F^p(U),
\end{equation}
so that $ \Phi^\ast $ sends p-forms into p-forms.
\newline
It is well known that the exterior derivative commutes with
pull-back, so that
\begin{equation}
\Phi^\ast(d\omega^{(\mathbf{A})}_{(\mathbf{B})})=d\Phi^\ast(\omega^{(\mathbf{A})}_{(\mathbf{B})}),
\end{equation}
and
\begin{equation}
\Phi^\ast(d\omega^{(\mathbf{A})})=d\Phi^\ast(\omega^{(\mathbf{A})}).
\end{equation}
We also have
\begin{equation}
\Phi^\ast(\omega^{(\mathbf{A})}_{(\mathbf{B})}\wedge\omega^{(\mathbf{B})})=
\Phi^\ast(\omega^{(\mathbf{A})}_{(\mathbf{B})})\wedge\Phi^\ast(\omega^{(\mathbf{B})}).
\end{equation}
The equation (2.10) can be seen as pull-back,
\begin{equation}
\Phi^\ast(\omega^{(\mathbf{A})})=dtz^{(\mathbf{A})}+\varpi^{(\mathbf{A})}.
\end{equation}
It can be shown, by a simple calculation that
\begin{equation}
\Phi^\ast(\omega^{(\mathbf{A})}_{(\mathbf{B})})=
\varpi^{(\mathbf{A})}_{(\mathbf{B})}.
\end{equation}
\newpage
By the exterior derivative of (2.22), we obtain
\begin{eqnarray}
\nonumber
d(\Phi^\ast(\omega^{(\mathbf{A})}))=d(dtz^{(\mathbf{A})}+\varpi^{(\mathbf{A})})=dz^{(\mathbf{A})}\wedge(dt)\\
\nonumber+dt\wedge\frac{\partial(\varpi^{(\mathbf{A})})}{\partial(t)}\\
\end{eqnarray}
+ terms not involving  $dt$.
\vspace{1cm}
\newline
Making a pull-back of (2.15) and using (2.20) we have
\begin{equation}
\Phi^\ast(d\omega^{(\mathbf{A})})=\Phi^\ast(-\omega^{(\mathbf{A})}_{(\mathbf{B})}\wedge\omega^{(\mathbf{B})})=
-\Phi^\ast(\omega^{(\mathbf{A})}_{(\mathbf{B})})\wedge\Phi^\ast(\omega^{(\mathbf{B})}).
\end{equation}
Using (2.19), (2.22), (2.23) and (2.24) we have
\begin{equation}
\frac{\partial(\varpi^{(\mathbf{A})})}{\partial(t)}=dz^{(\mathbf{A})}+
\varpi^{(\mathbf{A})}_{(\mathbf{B})}z^{(\mathbf{D})}.
\end{equation}
We can, by a similar procedure to (2.19), and using the Cartan's
second structure equation, obtain the following result
\begin{equation}
\frac{\partial(\varpi_{(\mathbf{A})(\mathbf{B})})}{\partial(t)}=
R_{(\mathbf{A})(\mathbf{B})(\mathbf{C})(\mathbf{D})}z^{(\mathbf{C})}\varpi^{(\mathbf{A})}.
\end{equation}
Making a new partial derivative of (2.25), two partial derivatives
of (2.9), comparing the results and using (2.26) we have the
following equation
\begin{equation}
\frac{\partial^2(A_{(\mathbf{A}){(\mathbf{C})(\mathbf{D})}})}{\partial(t^2)}=
tz^{(\mathbf{B})}R_{(\mathbf{A})(\mathbf{B})(\mathbf{C})(\mathbf{D})}+
z^{(\mathbf{L})}z^{(\mathbf{M})}R_{(\mathbf{A})(\mathbf{L})(\mathbf{M})(\mathbf{N})}
A_{(\mathbf{P}){(\mathbf{C})(\mathbf{D})}}\eta^{(\mathbf{N})(\mathbf{P})}.
\end{equation}
It is easy to show that
\begin{equation}
A_{(\mathbf{A}){(\mathbf{C})(\mathbf{D})}}+A_{(\mathbf{A}){(\mathbf{D})(\mathbf{C})}}=0
\end{equation}
is the solution for all t.
\newline
Then,
\begin{equation}
A_{(\mathbf{A}){(\mathbf{C})(\mathbf{D})}}=-A_{(\mathbf{A}){(\mathbf{D})(\mathbf{C})}},
\end{equation}
so that,  we can rewrite (2.9) as
\begin{equation}
\varpi^{(\mathbf{A})}= tdz^{(\mathbf{A})}+
\frac{1}{2}A^{({A})_{(\mathbf{B})(\mathbf{C})}}(z^{(\mathbf{B})}dz^{(\mathbf{C})}-z^{(\mathbf{C})}dz^{(\mathbf{B})}).
\end{equation}
Let us  define
\begin{equation}
A_{(\mathbf{A}){(\mathbf{C})(\mathbf{D})}}=z^{(\mathbf{B})}B_{(\mathbf{A}){(\mathbf{B})(\mathbf{C})(\mathbf{D})}}.
\end{equation}
The following result is obtained by substituting (2.31) in (2.27),
\begin{equation}
\frac{\partial^2(B_{(\mathbf{A}){(\mathbf{B})(\mathbf{C})(\mathbf{D})}})}{\partial(t^2)}=
tR_{(\mathbf{A})(\mathbf{B})(\mathbf{C})(\mathbf{D})}+
z^{(\mathbf{L})}z^{(\mathbf{M})}R_{(\mathbf{A})(\mathbf{B})(\mathbf{L})(\mathbf{N})}
B_{(\mathbf{P}){(\mathbf{M})(\mathbf{C})(\mathbf{D})}}\eta^{(\mathbf{N})(\mathbf{P})}.
\end{equation}
By a simple procedure we obtain the following solution
\begin{equation}
B_{(\mathbf{A}){(\mathbf{B})(\mathbf{C})(\mathbf{D})}}+B_{(\mathbf{B}){(\mathbf{A})(\mathbf{C})(\mathbf{D})}}=0.
\end{equation}
Using (2.29), (2.31) and (2.33) we conclude that
$B_{(\mathbf{A}){(\mathbf{B})(\mathbf{C})(\mathbf{D})}} $ has the
same symmetries as the Riemann curvature tensor
\begin{equation}
B_{(\mathbf{A}){(\mathbf{B})(\mathbf{C})(\mathbf{D})}}=-B_{(\mathbf{B}){(\mathbf{A})(\mathbf{C})(\mathbf{D})}}=
-B_{(\mathbf{A}){(\mathbf{B})(\mathbf{D})(\mathbf{C})}}.
\end{equation}
 Using (2.29) and  (2.31) we have
\begin{eqnarray}
\nonumber A_{(\mathbf{A}){(\mathbf{C})(\mathbf{D})}}dz^{(\mathbf{A})}z^{(\mathbf{C})}dz^{(\mathbf{D})}=\\
\nonumber +\frac{1}{4}B_{(\mathbf{A}){(\mathbf{B})(\mathbf{C})(\mathbf{D})}}.\\
\nonumber .(z^{(\mathbf{B})}dz^{(\mathbf{A})}-z^{(\mathbf{A})}dz^{(\mathbf{B})}).\\
\nonumber .(z^{(\mathbf{C})}dz^{(\mathbf{D})}-z^{(\mathbf{D})}dz^{(\mathbf{C})}).\\
\end{eqnarray}
Now we can write the line element of the hyper-surface. We have
\begin{eqnarray}
\nonumber ds'^2=t^2\eta_{(\mathbf{A})(\mathbf{B})}dz^{(\mathbf{A})}dz^{(\mathbf{B})}+\\
 \nonumber+\frac{1}{2}\{\frac{1}{2}t\epsilon_{(\mathbf{B})}B_{(\mathbf{A}){(\mathbf{B})(\mathbf{C})(\mathbf{D})}}+\\
 \nonumber+\eta^{(\mathbf{M})(\mathbf{N})}A_{(\mathbf{M}){(\mathbf{B})(\mathbf{A})}}A_{(\mathbf{N}){(\mathbf{C})(\mathbf{D})}}\}.\\
\nonumber.(z^{(\mathbf{B})}dz^{(\mathbf{A})}-z^{(\mathbf{A})}dz^{(\mathbf{B})})(z^{(\mathbf{C})}dz^{(\mathbf{D})}-z^{(\mathbf{D})}dz^{(\mathbf{C})}).\\
\end{eqnarray}
\newpage
The line elements of the manifold and the hyper-surface are equal
at $ t=1 $, where  $u^{\Lambda}=v^{\Lambda} $,
\begin{equation}
ds^2=ds'^2,
\end{equation}
and
\begin{eqnarray}
\nonumber ds^2=\eta_{(\mathbf{A})(\mathbf{B})}dz^{(\mathbf{A})}dz^{(\mathbf{B})}+\\
 \nonumber+\frac{1}{2}\{\frac{1}{2}\epsilon_{(\mathbf{B})}B_{(\mathbf{A}){(\mathbf{B})(\mathbf{C})(\mathbf{D})}}+\\
 \nonumber+\eta^{(\mathbf{M})(\mathbf{N})}A_{(\mathbf{M}){(\mathbf{B})(\mathbf{A})}}A_{(\mathbf{N}){(\mathbf{C})(\mathbf{D})}}\}.\\
\nonumber.(z^{(\mathbf{B})}dz^{(\mathbf{A})}-z^{(\mathbf{A})}dz^{(\mathbf{B})})(z^{(\mathbf{C})}dz^{(\mathbf{D})}-z^{(\mathbf{D})}dz^{(\mathbf{C})}).\\
\end{eqnarray}
In the next section we build, by a simple procedure, the conformal
form of n-dimensional pseudo-Riemannian manifolds.
\renewcommand{\theequation}{\thesection.\arabic{equation}}
\section{\bf Conformal Form of Riemannian Metrics }
 \setcounter{equation}{0}
$         $
 \setcounter{equation}{0}
 $         $
We now write (2.38) as
\begin{eqnarray}
 \nonumber ds^2=\eta_{(\mathbf{A})(\mathbf{B})}dz^{(\mathbf{A})}dz^{(\mathbf{B})}+\\
 \nonumber +\{\frac{1}{2}[\frac{1}{2}\epsilon_{(\mathbf{B})}B_{(\mathbf{A}){(\mathbf{B})(\mathbf{C})(\mathbf{D})}}+\\
 \nonumber +\eta^{(\mathbf{M})(\mathbf{N})}A_{(\mathbf{M}){(\mathbf{B})(\mathbf{A})}}A_{(\mathbf{N}){(\mathbf{C})(\mathbf{D})}}]\}.\\
\nonumber .(z^{(\mathbf{B})}\frac{dz^{(\mathbf{A})}}{ds}-z^{(\mathbf{A})}\frac{dz^{(\mathbf{B})}}{ds})(z^{(\mathbf{C})}\frac{dz^{(\mathbf{D})}}{ds}-z^{(\mathbf{D})}\frac{dz^{(\mathbf{C})}}{ds}))ds^2.\\
\end{eqnarray}
It can also  be written in the form
\begin{eqnarray}
 \nonumber[1-\frac{1}{2}[\frac{1}{2}\epsilon_{(\mathbf{B})}B_{(\mathbf{A}){(\mathbf{B})(\mathbf{C})(\mathbf{D})}}+\\
 \nonumber+\eta^{(\mathbf{M})(\mathbf{N})}A_{(\mathbf{M}){(\mathbf{B})(\mathbf{A})}}A_{(\mathbf{N}){(\mathbf{C})(\mathbf{D})}}].\\
\nonumber.(z^{(\mathbf{B})}\frac{dz^{(\mathbf{A})}}{ds}-z^{(\mathbf{A})}\frac{dz^{(\mathbf{B})}}{ds})(z^{(\mathbf{C})}\frac{dz^{(\mathbf{D})}}{ds}-z^{(\mathbf{D})}\frac{dz^{(\mathbf{C})}}{ds})]ds^2\\
 \nonumber =\eta_{(\mathbf{A})(\mathbf{B})}dz^{(\mathbf{A})}dz^{(\mathbf{B})}.\\
\end{eqnarray}
We now define the function
\begin{eqnarray}
L^{\mathbf{A})(\mathbf{B})}=(z^{(\mathbf{B})}\frac{dz^{(\mathbf{A})}}{ds}-z^{(\mathbf{A})}\frac{dz^{(\mathbf{B})}}{ds}),
\end{eqnarray}
which is the classical angular momentum  of a free particle.
\newline
The line element (3.2) can assume the following form
\begin{eqnarray}
 \nonumber\{1+\frac{1}{2}[\frac{1}{2}(\epsilon_{(\mathbf{B})}B_{(\mathbf{A}){(\mathbf{B})(\mathbf{C})(\mathbf{D})}}+\\
 \nonumber+\eta^{(\mathbf{M})(\mathbf{N})}A_{(\mathbf{M}){(\mathbf{B})(\mathbf{A})}}A_{(\mathbf{N}){(\mathbf{C})(\mathbf{D})}}].\\
\nonumber.(L^{\mathbf{A})(\mathbf{B})}L^{\mathbf{C})(\mathbf{D})})\}ds^2\\
 \nonumber =(\eta_{(\mathbf{A})(\mathbf{B})}dz^{(\mathbf{A})}dz^{(\mathbf{B})}.\\
\end{eqnarray}
We now define the function
\begin{eqnarray}
\nonumber \exp(-2\sigma)=\{1+\frac{1}{2}[\frac{1}{2}(\epsilon_{(\mathbf{B})}B_{(\mathbf{A}){(\mathbf{B})(\mathbf{C})(\mathbf{D})}}\\
 \nonumber +\eta^{(\mathbf{M})(\mathbf{N})}A_{(\mathbf{M}){(\mathbf{B})(\mathbf{A})}}A_{(\mathbf{N})){(\mathbf{C})(\mathbf{D})}})].\\
\nonumber .L^{(\mathbf{A})(\mathbf{B})}L^{(\mathbf{C})(\mathbf{D})}\},\\
\end{eqnarray}
so that, the line element assumes the form
\begin{equation}
ds^2=\exp(2\sigma)\eta_{(\mathbf{A})(\mathbf{B})}dz^{(\mathbf{A})}dz^{(\mathbf{B})}.
\end{equation}
The metric (3.6) is conformal to a flat manifold, and we conclude
that all n-dimensional pseudo-Riemannian metrics are conformal to
flat manifolds, when, in normal coordinates, the transformations are
well-behaved in the origin and in its neighborhood. It is important
to pay attention to the fact that a normal transformation and its
inverse are well-behaved in the region where geodesics  are not
mixed. Points where geodesics close or mix are known as conjugate
points of Jacobi's fields. Jacobi's fields can be used for this
purpose.
\newline
We can place (3.6) in the following form, \cite{1},
\begin{eqnarray}
\nonumber ds^2=\{1+\frac{1}{2}[\frac{1}{2}(\epsilon_{\beta}B_{\alpha{\beta\gamma\delta}})\\
 \nonumber +\eta^{\rho\sigma}A_{\rho{\alpha\beta}}A_{\sigma{\gamma\delta)}})].\\
\nonumber .L^{\alpha\beta}L^{\gamma\delta}\}^{-1}\eta_{\alpha\beta}d\Omega^{\alpha}d\Omega^{\beta}.\\
\end{eqnarray}
We now present the metric of a constant curvature manifold in the
well known form
\begin{eqnarray}
ds'^2=\{1+\frac{K\Omega^{\mathbf{\alpha}}\Omega^{\beta}\eta_{\alpha\beta}}{4}\}^{-2}d\Omega^{\rho}d\Omega^{\sigma}\eta_{\varrho\sigma}.
\end{eqnarray}
Because (3.7) and (3.8) are conformal to a flat manifold, there is a
conformal transformation between them,
\begin{eqnarray}
g'_{\alpha\beta}=(\exp2\psi)g_{\alpha\beta}.
\end{eqnarray}
More specifically,
\begin{eqnarray}
\nonumber\{1+\frac{1}{2}[\frac{1}{2}(\epsilon_{\beta}B_{\alpha{\beta\gamma\delta}})+\\
\nonumber
+\eta^{\rho\sigma}A_{\rho{\alpha\beta}}A_{\sigma{\gamma\delta)}})]L^{\alpha\beta}L^{\gamma\delta}\}=\\
 \nonumber
 =(\exp2\psi)\{1+\frac{K\Omega^{\mathbf{\alpha}}\Omega^{\beta}\eta_{\alpha\beta}}{4}\}^{2}.\\
 \end{eqnarray}
\newline
Note that (3.8) is an Einstein's space with a constant curvature,
where
\begin{eqnarray}
R'_{\alpha\beta}=\frac{R'}{n}g'_{\alpha\beta},
\end{eqnarray}
and $R'$ is the scalar curvature. Spaces, as the Schwarzschild's,
where
\begin{eqnarray}
R_{\alpha\beta}=0,
\end{eqnarray}
are  Einstein's spaces and are not maximally symmetric.
\newline
Einstein's spaces with a constant scalar curvature obey homogeneity
and isotropy conditions. They are maximally symmetric spaces.
\newline
We will be using the following definitions, \cite{5}
\begin{eqnarray}
\triangle_{1}{\psi}=g^{\mu\nu}{\psi}_{,\mu}{\psi}_{,\nu},
\end{eqnarray}
\begin{eqnarray}
{\psi}_{\mu\nu}={\psi}_{;\mu\nu}-{\psi}_{,\mu}{\psi}_{,\nu},
\end{eqnarray}
\begin{eqnarray}
\triangle_{2}{\psi}=g^{\mu\nu}{\psi}_{;\mu\nu}.
\end{eqnarray}
From (3.9), (3.13), (3.14), and (3.15) we obtain
\begin{eqnarray}
\nonumber {\psi}_{\mu\nu}=\frac{1}{(n-2)}(R_{\mu\nu})\\
\nonumber -\frac{1}{(2)(n-1)(n-2)}(g'_{\mu\nu}R'-g_{\mu\nu}R)\\
\nonumber-\frac{1}{2}\triangle_{1}{\psi}g_{\mu\nu}.\\
\end{eqnarray}
If $g'_{\mu\nu}$ is a metric of an Einstein's space, then (3.16) is
simplified to
\begin{eqnarray}
\nonumber {\psi}_{\mu\nu}=-\frac{1}{(n-2)}R_{\mu\nu}+\\
\nonumber +(\frac{1}{(2)(n-1)(n-2)}R+\frac{1}{(2n)(n-1)}R'(\exp2\psi)-\frac{1}{2}\triangle_{1}{\psi})g_{\mu\nu}.\\
\end{eqnarray}
In the following we consider the Einstein's equation
\begin{eqnarray}
\nonumber R_{\mu\nu}-\frac{1}{2}Rg_{\mu\nu}={8\pi}T_{\mu\nu}.\\
\end{eqnarray}
Spaces as (3.11) with a non-constant scalar curvature do not obey
(3.18).
\newpage
\section{Embedding of Manifolds of  Constant Curvatures in Flat Manifolds}
 $                       $
 \setcounter{equation}{0}
 $         $
In this section we embed  the n-dimensional manifold (3.8) in an
n+1-dimensional flat manifold obtaining, as a geometric result,
without postulate, the quantum angular momentum of a particle. Other
results will be presented in another section.
\newline
We now consider a manifold (3.8) called S, embedded in an
n+1-dimensional flat manifold. The following constraint is obeyed
\cite{6},
\begin{eqnarray}
\eta_{\alpha\beta}x^{\alpha}x^{\beta}=K=\epsilon\frac{1}{R^2},
\end{eqnarray}
where K is the  scalar curvature of the n-dimensional manifold
(3.8), $\alpha, \beta = ( 1, 2,...,n+1 )$ and $ \epsilon=(+1, -1).$
For the special case of an n-sphere we use the following notation
$S^n$ for (3.8).
\newline
It is convenient that we use a local basis
$X_{\beta}=\frac{\partial}{\partial(x^{\beta})}.$
\newline
We consider a constant vector $\textbf{C}$ in the n+1-dimensional
manifold given by
\begin{eqnarray}
\eta_{\alpha\beta}C^{\alpha}X^{\beta}=\eta^{\alpha\beta}C_{\alpha}X_{\beta}=C,
\end{eqnarray}
where each  $C^{\alpha}$ is constant and  $\textbf{N}$ is a unitary
and normal vector to S. We use the symbol $<,>$ for the internal
product in the n+1-dimensional flat manifold  and $<,>'$ for S.
\newline
A constant vector $\textbf{C}$ can be  decomposed into two parts,
one in S and the other outside S as follows
\begin{eqnarray}
C=\bar{C}+<C,N>N.
\end{eqnarray}
From the definition of $\textbf{N}$ and (4.1) we obtain
\begin{eqnarray}
N^{\alpha}=\frac{x^{\alpha}}{R}
\end{eqnarray}
Let us construct  the covariant derivative of $\textbf{C}$. We have
a local basis and  a diagonal and unitary tensor metric, so that the
Christoffel symbols are null. Then the covariant derivative of
$\textbf{C}$ in the $\textbf{Y}$ direction is given by
\begin{eqnarray}
\nabla_{Y}C=0.
\end{eqnarray}
It is easy to show that
\begin{eqnarray}
\nabla_{Y}N=\frac{Y}{R}.
\end{eqnarray}
The Lie derivative of the metric tensor in S is given by \cite{6},
\begin{eqnarray}
L_{\bar{U}}g'=2\lambda_{U}g',
\end{eqnarray}
where $\textbf{U}$ is a constant vector in the flat manifold, and
$\lambda_{U}$ is the  characteristic function. For S the
characteristic function is given by
\begin{eqnarray}
\lambda_{U}=-\frac{1}{R}<U,N>.
\end{eqnarray}
Substituting (4.8) in (4.7) we have
\begin{eqnarray}
L_{\bar{U}}g'=-2\frac{1}{R}<U,N>g'.
\end{eqnarray}
In the region of S where $<U,N>$ is not null,  $ \bar{U} $ is a
conformal Killing vector and in the region where  $<U,N>$  is null,
 $\bar{U} $ is a Killing vector.
\newline
We now consider another constant vector $\textbf{V}$ in the flat
space. The Lie derivative of its projection in S is given by
\begin{eqnarray}
L_{\bar{U}}g'=-2\frac{1}{R}<U,N>g'.
\end{eqnarray}
As we consider a local basis and constant vectors $\textbf{U}$ and
$\textbf{V}$, the commutator is given by
\begin{eqnarray}
[U,V]=0.
\end{eqnarray}
Then,
\begin{eqnarray}
L_{[\bar{U},\bar{V}]}g'=-2\frac{1}{R}<[U,V],N>g'=0.
\end{eqnarray}
Regardless of $ \bar{U}$ and $\bar{V}$ being conformal Killing
vectors or Killing vectors, their commutator is a Killing vector. In
the following we will show that the commutator $[\bar{U},\bar{V}]$
is proportional to the quantum angular momentum of a particle.
\newline
Using (3.3) in the following commutator of elements of the basis, we
obtain
\begin{eqnarray}
\nonumber [\bar{U},\bar{V}]=\\
\nonumber =U^{\alpha}V^{\beta}[X_\alpha-<X_\alpha,N>N,X_\beta-<X_\beta,N>N]=\\
\nonumber =U^{\alpha}V^{\beta}[\bar{X}_\alpha,\bar{X}_\beta].\\
\end{eqnarray}
We now calculate the commutator of elements of the basis, by parts.
\newline
 We have by simple calculation
\begin{eqnarray}
<X_\alpha,N>N=\frac{1}{R}\eta_{\alpha\beta}x^{\beta}.
\end{eqnarray}
Substituting (4.14) in (4.13) we obtain
\begin{eqnarray}
\nonumber [\bar{X}_\alpha,\bar{X}_\beta]=\\
\nonumber =[X_\alpha,X_\beta]-[X_\alpha,\frac{1}{R}\eta_{\beta\sigma}x^{\sigma}N]+[X_\beta,\frac{1}{R}\eta_{\alpha\sigma}x^{\sigma}N]+\\
\nonumber +\frac{1}{R^2}[\eta_{\alpha\sigma}x^{\sigma}N, \eta_{\beta\sigma}x^{\sigma}N].\\
\end{eqnarray}
In a local basis we have
\begin{eqnarray}
[X_\alpha,X_\beta]=0,
\end{eqnarray}
\begin{eqnarray}
[\eta_{\alpha\sigma}x^{\sigma}N, \eta_{\beta\sigma}x^{\sigma}N]=0.
\end{eqnarray}
Substituting in (4.15) we obtain
\begin{eqnarray}
\nonumber [\bar{X}_\alpha,\bar{X}_\beta]=\\
\nonumber =
\frac{1}{R^2}(\eta_{\alpha\sigma}x^{\sigma}\frac{\partial}{\partial(x^{\beta})}-\eta_{\beta\sigma}x^{\sigma}\frac{\partial}{\partial(x^{\alpha})})\\
\nonumber =
\frac{1}{R^2}(x_\alpha\frac{\partial}{\partial(x^{\beta})}-x_\beta\frac{\partial}{\partial(x^{\alpha})})\\
\nonumber=-i\frac{1}{\hbar}\frac{1}{R^2}L_{\alpha\beta}.\\
\end{eqnarray}
Multiplying $ L_{\alpha\beta}$ by a vielbein basis  we obtain
\begin{eqnarray}
\nonumber
L_{(\mathbf{A})(\mathbf{B})}=\\
\nonumber =(i\hbar)(R^{2})R_{(\mathbf{A})(\mathbf{B})(\mathbf{C})(\mathbf{D})}x^{(\mathbf{D})}\eta^{(\mathbf{C})(\mathbf{M})}\frac{\partial}{\partial(x^{\mathbf{M})}}.\\
\end{eqnarray}
where
\begin{eqnarray}
 \hat{p}_{(\mathbf{M})}=(i\hbar)\frac{\partial}{\partial(x^{\mathbf{M})}},
\end{eqnarray}
is the quantum momentum operator of a particle, and
\begin{eqnarray}
\nonumber
R_{(\mathbf{A})(\mathbf{B})(\mathbf{C})(\mathbf{D})}=\\
\nonumber
=\frac{1}{R^2}[\eta_{(\mathbf{A})(\mathbf{D})}\eta_{(\mathbf{B})(\mathbf{C})}-\eta_{(\mathbf{A})(\mathbf{C})}\eta_{(\mathbf{B})(\mathbf{D})}],\\
\end{eqnarray}
is the curvature of S in the vielbein basis  and $
\eta_{(\mathbf{A})(\mathbf{C})} $ is diagonal. We consider as an
important observation that the association between the quantum
angular momentum operator and the constant curvature operator is
allowed in an orthogonal vielbein basis of a Cartesian coordinate,
regardless of having a curved or a flat manifold. We have used the
embedding of an n-dimensional manifold S in an n+1-dimensional flat
manifold, only to obtain the quantum angular momentum operator of a
particle, without postulates. We can rewrite (4.19) as follows
\begin{eqnarray}
\nonumber
L_{(\mathbf{A})(\mathbf{B})}=\\
\nonumber
=(i\hbar)[\eta_{(\mathbf{A})(\mathbf{D})}\eta_{(\mathbf{B})(\mathbf{C})}-\eta_{(\mathbf{A})(\mathbf{C})}\eta_{(\mathbf{B})(\mathbf{D})}].\\
\nonumber
.x^{(\mathbf{D})}\eta^{(\mathbf{C})(\mathbf{M})}\frac{\partial}{\partial(x^{\mathbf{M})}}.\\
\end{eqnarray}
Note that the coordinates in (3.7) are in the n+1-dimensional flat
manifold  and  $L_{\alpha\beta}\subset S$, so that
$L_{\alpha\beta}=0$  for $\alpha$ or $\beta$ is equal to $n+1.$
\newline
Racah has shown that \cite{7} the Casimir operators of  any
semisimple Lie group can be constructed from the quantum angular
momentum (4.22). Each multiplet of semisimple Lie group can be
uniquely characterized by the eigenvalues  of the Casimir operators.
\newline
We have built the quantum angular momentum operator from classical
geometric considerations. We can write the usual expression for the
eigenstates of Casimir operator without reference to quantum
mechanics, as follows
\begin{eqnarray}
\hat{C}\mid...>=C\mid...>.
\end{eqnarray}
This is simple and well known,\cite{7} for SO(3). There is an
interesting construction from the group theoretical point of view to
Dirac theory with or without Dirac's equation, \cite{8}.
\newline
In the following we calculate the Lie derivative of the so(p,n-p)
algebra. For the Lie group SO(p,n-p) we choose the signature
\newline
$(p,n-p)=(-,-,-,...-,+,+,..+)$, with the algebra
\begin{eqnarray}
\nonumber
[L_{(\mathbf{A})(\mathbf{B})},L_{(\mathbf{C})(\mathbf{D})}]=-i(\eta_{(\mathbf{A})(\mathbf{C})}L_{(\mathbf{B})(\mathbf{D})}+\eta_{(\mathbf{A})(\mathbf{D})}L_{(\mathbf{C})(\mathbf{B})}\\
\nonumber +\eta_{(\mathbf{B})(\mathbf{C})}L_{(\mathbf{D})(\mathbf{A})}+\eta_{(\mathbf{B})(\mathbf{D})}L_{(\mathbf{A})(\mathbf{C})}).\\
\end{eqnarray}
Considering the Lie derivative
\begin{eqnarray}
\nonumber \textbf{L}_{[L_{(\mathbf{A})(\mathbf{B})},L_{(\mathbf{C})(\mathbf{D})}]}g'=\\
\nonumber
=-R^{4}<[[X_{(\mathbf{A})},X_{(\mathbf{B})}],[X_{(\mathbf{C})},X_{(\mathbf{D})}]],N>g'=0,\\
\end{eqnarray}
where, for the orthogonal Cartesian coordinates, the vielbein is
given by
\begin{equation}
E_{\Lambda}^{(\mathbf{A})}=\delta_{\Lambda}^{(\mathbf{A})},
\end{equation}
we have
\begin{eqnarray}
\nonumber[X_{(\mathbf{A})},X_{(\mathbf{B})}]=[X_{\alpha},X_{\beta}]=0.\\
\end{eqnarray}
\newline
Note that $g'$ in S is form-invariant \cite{9} in relation to the
Killing's vector $\xi$, and in relation to the  algebra of
SO(p,n-p), as well. We conclude that the algebra of SO(p,n-p)  is a
Killing's object.
\newpage
The same is true for the algebra of the Lie group SO(n), where for
SO(n) we could  choose the following signature $(+,+,+...,+,+)$. The
constraint (4.1) is invariant for many of the classical groups. For
these groups it is possible to build operators, from the combination
of the quantum angular momentum operators, which are Killing's
objects in relation to $´g'$. Therefore, the metric is
form-invariant in relation to this algebra. It is interesting to see
some of these groups in the Cartan's list of irreducible Riemannian
globally symmetric spaces, \cite{4}, and in \cite{10}.
\newline
Note that we start from a normal coordinate transformation. In other
words, in the region where the transformation (2.4) is well-behaved,
we can build (3.6) and by a conformal transformation we have (3.8),
which was essential to obtain the quantum angular momentum operator
from geometry.
\newpage
\begin{center}
\section{Physical Principles Based on Geometric Properties}
 $                       $
\end{center}
 \setcounter{equation}{0}
 $         $
From a different point of view, Dirac \cite{12} embedded the De
Sitter space in a five-dimensional flat manifold. He has considered
functions and fields living in a five-dimensional flat manifold and
has built a procedure to project them in the De Sitter space. The
Dirac procedure implies  the need for the quantum momentum and the
quantum angular momentum postulates. Other authors have used the
Dirac's idea or variants of it, as in \cite{13}. To obtain the
quantum angular momentum from geometric considerations, we have
considered only constant vectors in an n+1-dimensional flat
manifold. There are many procedures to define or introduce
functions, fields and geometric objects in (3.8). In the following,
we reconsider the qualitative analyzes of (4.9) made in section $4$.
In the region where $<U,N>$ does not vanish  $\bar{U}$ is a
conformal Killing vector and in the region where it vanishes,
$\bar{U}$ is a Killing vector. In other words, we have Killing and
conformal Killing vectors living in (3.8). For our objective we need
only Killing objects as the quantum angular momentum.
\newline
If we make the Lie derivative of (3.9) in relation to $\xi$, we
obtain
\begin{eqnarray}
L_{\xi}g'=(2\xi(\psi)g+L_{\xi}g)(\exp2\psi).
\end{eqnarray}
More specifically, (5.1) can be written as
\begin{eqnarray}
\nonumber
 L_{\xi}[\{1+\frac{K\Omega^{\mathbf{\alpha}}\Omega^{\beta}\eta_{\alpha\beta}}{4}\}^{(-2)}\eta]=\\
\nonumber L_{\xi}[(\exp2\psi)\{1+\frac{1}{2}[\frac{1}{2}(\epsilon_{\beta}B_{\alpha{\beta\gamma\delta}})+\\
\nonumber
+\eta^{\rho\sigma}A_{\rho{\alpha\beta}}A_{\sigma{\gamma\delta)}}]L^{\alpha\beta}L^{\gamma\delta}\}^{(-1)}\eta].\\
\end{eqnarray}
We now consider the following condition
\begin{eqnarray}
2\xi(\psi)g+L_{\xi}g=0,
\end{eqnarray}
which implies the following
\begin{eqnarray}
L_{\xi}g'=0,
\end{eqnarray}
which is a definition of a Killing vector. If  $\xi$ obeys (5.3) we
conclude that $\xi$ is a  Killing vector in (3.8). Note that $\xi$
is a conformal Killing vector in (3.7). The equation (5.3) shows how
a Killing vector in (3.8) will be in (3.7). We conclude that the
best way toward our objective will be from a postulate as follows,
associated to conditions of minimal energy : \textbf{In (3.8) nature
always choose Killing objects.}. Based on this postulate we will
build four classical principles, where one of them is identified as
a classical version of the Heisemberg's uncertainty principle and
another as a classical version of the Bohr's non-radiation
postulate. The third principle is not new and is associated to the
electric neutrality of some stable systems. The fourth  can be
interpreted as an equivalence between two descriptions of a
particle's motion. The first one as the motion due to the presence
of forces and the second as a consequence of geometry, as in
Einstein's gravitation. For this we assume only constant vectors in
an n+1-dimensional flat manifold, where (3.8) will be embedded.
\newline
The equations (2.27) and (2.32) tell us that if the curvature is
null, $ A_{(\mathbf{A}){(\mathbf{C})(\mathbf{D})}}$ and
$B_{(\mathbf{A}){(\mathbf{B})(\mathbf{C})(\mathbf{D})}}$ are null.
In this case, the equation (3.4) implies a null angular momentum. We
conclude that any free particle in a curved manifold will be always
in movement, with angular momentum not null regardless of wether or
not we consider a physical theory.
\newline
The equation (3.6) tells us that $ ds^2$ is conformal to a flat
manifold and to (3.8). An observer in (3.8) will see the space as
being homogeneous and isotropic in the small region where the
transformation (2.4) is well-behaved. With this condition, Ricci
principal directions of space will be indeterminate so that in that
region the position of the particle is uncertain. In the conjugate
points of Jacobi's fields, the transformation (2.4) fails because
geodesics cross, mix or touch each other. Therefore, close to a
conjugate point we will not have indetermination in the Ricci
principal directions and the uncertainty in the position of the
particle disappears. If (3.8) is valid in all points of the space,
in each point there will be an indetermination of Ricci principal
directions and consequently a total uncertainty in the position of
the particle. This resembles a property of the Heisenberg's
uncertainty principle and could be seen as a classical version.
\newline
The metric (3.8) will be form-invariant for  a displacement $ \xi $
which is a Killing's vector. In this metric \cite{9}, a scalar
function will be constant or null, there will be Killing's vectors
only, and tensors will be a combination of the metric tensor. In
these conditions, the electromagnetic fields will be trivial and
there will not be radiation. In the neighborhoods of the conjugate
points the transformations in normal coordinates fail and we will
not have an indetermination of Ricci principal directions and the
electromagnetic fields will not be trivial, being a radiative field.
This is similar to the Bohr's postulate for radiation and could be
seen as a classical version.
\newline
In the metric (3.8) there are no forces generated by fields in the
region where the transformation (2.4) is well-behaved. Particles
move free of forces. In the local system, there are ordinary  forces
generated by fields. If we consider a Kalusa-Klein theory, where
gauge fields are present in the metric, particles are free in the
local coordinates and in (3.6). In the local system, there are
fields while in (3.8) there are not.
\newline
We can consider this as a principle, creating  an equivalence
between two descriptions of motion, which are possible by normal
transformations. The first description, in local coordinates, is
related to the conception of force generating fields. The second is
related to the conception of  motion without forces.
\newline
We believe that this principle is going toward the  Einstein's dream
because it points to  the possibility of thinking in  physics
without forces as in Einstein's gravity.
\newline
We notice that the conjugate points of the Jacobi's fields can be a
consequence of  geometric singularities, as it is in the origin of
the Schwarzschild's geometry, \cite{11}, where the curvature
diverges, but it can also be due to the construction of the
coordinates, as it is in the case of a maximally symmetric space,
where the curvature is finite in all points. In the second case we
have an indetermination of Ricci principal directions, and in the
first we do not.
\newline
Considering the momentum-energy tensor of matter and electromagnetic
fields,
\begin{eqnarray}
\nonumber T_{\alpha\beta}=\frac{1}{4\pi}(F_{\alpha\sigma}F_{\beta}{^{\sigma}}-\frac{1}{4}g'_{\alpha\beta}F_{\varrho\sigma}F^{\rho\sigma})+t_{\alpha\beta},\\
\end{eqnarray}
where $ t_{\alpha\beta} $ is associated to  electric charges and
$T(F_{\alpha\sigma})=T_{\alpha\beta\mathbf{em}}$ to the
electromagnetic fields, with
\begin{eqnarray}
\nonumber T_{\alpha\beta\mathbf{em}}=\frac{1}{4\pi}(F_{\alpha\sigma}F_{\beta}{^{\sigma}}-\frac{1}{4}g'_{\alpha\beta}F_{\varrho\sigma}F^{\rho\sigma}).\\
\end{eqnarray}
Then,
\begin{eqnarray}
\nonumber T_{\alpha\beta}=T_{\alpha\beta\mathbf{em}}+t_{\alpha\beta}.\\
\end{eqnarray}
As  $g'$ is form-invariant in (3.8), the electromagnetic vector $
A_{\mu}$ will be null in (3.8),
\begin{eqnarray}
A_{\mu}=0.
\end{eqnarray}
\newline
From Maxwell's equations we have
\begin{eqnarray}
F^{\rho\sigma}{_{;\rho}}=-J^{\sigma}.
\end{eqnarray}
Using(5.9) in (5.10) we obtain
\begin{eqnarray}
 J^{\sigma}=0.
\end{eqnarray}
We conclude that in (3.8) the sum of all charges is zero, as well as
it is the sum of all currents.
\begin{center}
\newpage
\section{Geometric Properties Based on Quantum Principles and Quantum Principles Based on Geometric Properties }
 $                       $
\end{center}
 \setcounter{equation}{0}
 $         $
In this section we analyze some results obtained in section 5 which
resemble some postulates of quantum mechanics. Part of the
development of our results is qualitative because only some
applications, like SO(3) and the Casimir  eigenvalues, for instance,
can be easily calculated. Considerations of more complex systems are
qualitative at the moment. Therefore, from a theoretical point of
view, there is a gap.
\newline
We recall that in the region where there are no  conjugate points of
Jacobi's fields, it is possible to build a transformation (2.4)
between the ordinary metric and (3.6), and a conformal
transformation between (3.6) and (3.8). Because $g'$ is
form-invariant in the region where (3.8) is defined, there are no
fields nor radiation. The quantum angular momentum, which  is a
Killing's object, appears as a geometric consequence of embedding
(3.8) in an n+1-dimensional flat manifold. Particles will be in a
free motion, but confined in (3.8). In this context, where forces do
not exist, the particle confinement is due to the manifold geometry.
This resembles the Heisenberg's principle of quantum mechanics.
\newline
We consider as an important observation, made in section $5$, that
the association between the quantum angular momentum operator and
the constant curvature operator is allowed in a orthogonal vielbein
basis of a Cartesian coordinate, even for a flat manifold. This
suggests that, even in a flat spacetime, we can consider the
intrinsic angular momentum, or spin $\frac{1}{2}$ of a free massive
particle, as a quantum object living in an manifold of constant
curvature, embedded in this flat spacetime.
\newline
From the geometric point of view, some operations with the quantum
angular momentum, as sums and products, suggest the same operations
with curvature. We can define some procedures in differential
geometry by operations with quantum angular momenta. As an example,
consider the algebra (4.24) of the group SO(p,n-p)
\begin{eqnarray}
\nonumber
[L_{(\mathbf{A})(\mathbf{B})},L_{(\mathbf{C})(\mathbf{D})}]=-i(\eta_{(\mathbf{A})(\mathbf{C})}L_{(\mathbf{B})(\mathbf{D})}+\eta_{(\mathbf{A})(\mathbf{D})}L_{(\mathbf{C})(\mathbf{B})}\\
\nonumber +\eta_{(\mathbf{B})(\mathbf{C})}L_{(\mathbf{D})(\mathbf{A})}+\eta_{(\mathbf{B})(\mathbf{D})}L_{(\mathbf{A})(\mathbf{C})}).\\
\end{eqnarray}
We can substitute  (4.19) in (6.1) obtaining a representation of the
algebra in terms of  curvature operators. Substituting (4.19) in
(4.25) we will have the form-invariance of the metric tensor $g'$,
in relation to the algebra  so(p,n-p), in terms of the curvature
operators. Any other possible operation among quantum angular
momenta, here defined, can be placed in terms of curvatures. This
offers some curious procedures in differential geometry by simple
operations with quantum angular momentum, which may not be possible
by geometric methods. We do not know if this is known in the
specialized literature. The association between the quantum angular
momentum and differential geometry can be useful in geometry, as
well as in physics.
\newline
In the following, we make more qualitative considerations. We know
that mass and energy curve the spacetime. From the postulate of
section $5$, the existence of electric charges is allowed in (3.8)
provided that the total sum is zero. For each charge there is an
associated particle . The particles confinement in the metric (3.8)
is not due to forces, it is a consequence of the form-invariance of
$g'$. Obviously, in the region where there is a transition between
well-behaved and not well-behaved transformations (2.4), we also
have a transition from the condition where there are no forces to an
ordinary description by forces. This can be seen as a small
deformation in (3.8). The intensity of the deformation could be
responsible for some polarization or emission.
\newline
For a stable and isolated  system, like SO(3), we can write the
eigenvalues of the Casimir operators by (4.23). Particles will be
confined if the metric is (3.8). As we have seen, the postulate
implies the electric neutrality of (3.8) and  the confinement of the
particles.  We imagine that an incident particle curves and deforms
(3.8), then, there will be a transition from the well-behaved to the
not well-behaved transformation (2.4). This can be seen as a
transition, emission or scattering. Mass and energy of each particle
curve the spacetime, then particles in (3.8) will be responsible for
the confinement. We conclude that each particle contributes to the
confinement, which is proportional to its mass if the resultant
metric is (3.8). In this case a proton curves the spacetime with
more intensity than an electron because its mass is bigger. As a
consequence, the associated curvatures obey
\begin{eqnarray}
K_{\mathbf{p}}>K_{\mathbf{e}}.
\end{eqnarray}
But
\begin{eqnarray}
K=\frac{1}{R^2},
\end{eqnarray}
then
\begin{eqnarray}
R_{\mathbf{p}}<R_{\mathbf{e}}.
\end{eqnarray}
Note that the region of the electron  confinement is bigger than
that of the proton. This is compatible with the Heisemberg's
uncertainty principle and with the experimental evidence of nuclear
and atomic dimensions. Protons live in  nucleus and electrons live
outside. Starting from the above point of view, if the metric
deformation, caused by mass, energy and motion of some interacting
particles, generates a resultant metric given by (3.8), there will
not be any transition, emission, or scattering. If the metric
deformation generates a different metric from (3.8), there will be
transition, emission or scattering. This resembles the Bohr's
principle of quantum mechanics. In this paper we are using the
classical point of view and Riemannian manifolds. Considering an
ionized positive system, we conclude that the metric will not be
(3.8) because the sum of all charges is zero only for (3.8).
Although, the Einstein's equation is well-behaved for an ionized
system, we know that it is not valid for small regions. Obviously,
in the limit of a classical distribution of charges, the Einstein's
equation  is a good theory, with a negative curvature in the
interior of a classical distribution of charges, and null curvature
outside, obeying the weak energy condition, \cite{14}. For coherence
with this classical limit, we consider that a microscopic ionized
positive system has a negative curvature, so that the Jacobi
equation in a Fermi-Walker transported vielbein basis  will have a
form similar to an inverted oscillator, \cite{15}. A positive
charged incident particle will have in this space a trajectory which
will be interpreted as a repulsive electric force. If we have a
negative charged incident particle interacting with an ionized
positive system, the sum of all charges will be zero and by the
postulate above presented, another system with positive curvature
will be in process. The conditions of minimal energy of (3.8) will
be obeyed, resulting in energy loss (fotons) and a new electrically
neutral system (3.8) will be formed. This process will be seen as an
attractive electric force. If the new system is not electrically
neutral, therefore having a net positive charge, internal parts of
the system will be electrically neutral, shielding parts of the
system. The resultant geometry can be very complex. It will be
impossible to understand geometric details only by geometric
methods. However, the appropriate and usual quantum mechanics
operations with angular momenta can be converted in geometric
operations as we have seen in the beginning of this section. This
quantum mechanics  approach to geometry, associated to the
Landau-Raychaudhuri equation \cite{14} can be very useful.
\newline
The Heisemberg's and the Bohr's postulates are part of a theory
involving  force and interaction, known as quantum mechanics. We
have not exchanged the Heisemberg's postulate and the Bohr's
postulate by geometric properties of classical nature. Actually, we
have made classical geometric descriptions with  similar properties,
as much as possible, to the quantum postulates. We offered the
possibility of obtaining a quantum mechanics without forces, from
only one postulate , simple systems, and geometry. From our point of
view this theory will preserve the essence of quantum mechanics,
differing from the usual one in some aspects above presented . This
competitive classical geometric theory, if possible, will avoid the
incompatibilities between quantum mechanics and relativity,
verifying the success of both. This would represent a program of
very unexpected results and would not be attractive enough to
researchers because forces are considered as spacetime deformations.
Although we have presented a geometric alternative to electric
forces, unfortunately this is an unsolvable problem at the moment,
so that force will be present in many situations. The conceptions of
force and non-force may be seen, provisionally, as complementary. We
believe that our results can be considered as a possible starting
point to a quantum theory without forces.


\newpage
\begin{thebibliography}{99}
\bibitem{1}A.C.V.V.de Siqueira, {\it arXiv:math-ph/ 1006.2868v1}
\bibitem{2}E.Cartan,{\bf Le\c{c}\~ons sur la Geometrie des Espaces De Riemann},
(Gauthier-Villars, Paris,1946).
\bibitem{3}M.Spivak, {\bf A Comprehensive Introduction to Diferential Geometry}, {Volume two},
(Publish or Perish, Inc. 1999).
\bibitem{4}S.Helgason, {\bf Differential Geometry and Symmetric Spaces}
(Academic Press,1962)
\bibitem{5}L.P.Eisenart, {\bf Riemannian Geometry}
(Princeton University Press,1997)
\bibitem{6}W.C.Weber and S.I.Goldberg,{\it Queen's Papers in Pure and Applied Mathematics-No.16}
(Queen's University.Kingston.Ontaro,1969).
\bibitem{7}W.Greiner and B.Muller, {\bf Quantum Mechanics Symmetries}
(Springer,1994).
\bibitem{8}A.O.Barut {\it Phys. Rev. Letters}(Vol.20,893,1968)
\bibitem{9} S.Weinberg,{\bf Gravitation and Cosmology:Principles and Applications of the General Theory of
Relativity} (John Wiley Sons1972)
\bibitem{10}R.Gilmore, {\bf Lie Groups,Lie Algebras, and Some of Their
Applications} (Dover Publications,Inc,2002)
\bibitem{11} C. W. Misner,K. S. Thorne and J. A. Wheeler, {\bf Gravitation} (W. H. Freeman and
Company, San Francisco, 1973).
\bibitem{12} P.A.M.Dirac,{\it Ann.Math.} (vol36,657,1935).
\bibitem{13} R.Banerjee,{\it Ann.Phys.} (311,245,2004).
\bibitem{14}  S. W. Hawking and G. F. R. Ellis, {\bf The Large Scale
Structure of Space-Time} (Cambridge University Press,
Cambridge,1973)
\bibitem{15}See eq.(5.8) in  A. C. V. V. de Siqueira, {\it arXiv:math-ph/0802.2299v1}
\end{thebibliography}
\end{document}